\documentclass[conference, letterpaper]{IEEEtran}
\usepackage[utf8]{inputenc} 
\usepackage[T1]{fontenc}
\usepackage{url}
\usepackage{ifthen}
\usepackage{titlesec}
\usepackage[noadjust]{cite}
\usepackage[cmex10]{amsmath} 
\usepackage{amssymb}
\usepackage{graphicx}
\usepackage{epstopdf}
\usepackage{epsfig}
\usepackage{mathtools}

\usepackage{multirow}
\usepackage{array}
\usepackage{makecell}
\usepackage{tikz}
\usepackage{diagbox}
\usepackage{adjustbox}
\usepackage{caption, multirow, makecell}
\usepackage{blkarray}
\usepackage{nicematrix}
\usepackage{booktabs}
\usepackage{mathtools}
\usepackage{csquotes}
\usepackage{algorithm}
\usepackage{algorithmic}
\usepackage{comment}
\usepackage[english]{babel}

\newtheorem{thm}{Theorem}
\newtheorem{lem}{Lemma}

\newtheorem{corollary}{Corollary}

\newtheorem{defn}{Definition}

\newtheorem{exmp}{Example}

\setcellgapes{3pt}

\newtheorem{rem}{Remark}

\begin{document}

\title{Two-Dimensional Multi-Access Coded Caching with Multiple Transmit Antennas
}

\author{\IEEEauthorblockN{K. K. Krishnan Namboodiri, Elizabath Peter, and B. Sundar Rajan}
\IEEEauthorblockA{{Department of Electrical Communication Engineering, Indian Institute of Science, Bengaluru, India} \\
\{krishnank, elizabathp, bsrajan\}@iisc.ac.in}
}

\maketitle

\begin{abstract}	
This work introduces a multi-antenna coded caching problem in a two-dimensional multi-access network, where a server with $L$ transmit antennas and $N$ files communicates to $K_1K_2$ users, each with a single receive antenna, through a wireless broadcast link. The network consists of $K_1K_2$ cache nodes and $K_1K_2$ users. The cache nodes, each with capacity $M$, are placed on a rectangular grid with $K_1$ rows and $K_2$ columns, and the users are placed regularly on the square grid such that a user can access $r^2$ neighbouring caches in a cyclic wrap-around fashion. For a given cache memory $M$, the goal of the coded caching problem is to serve the user demands with a minimum delivery time. We propose a solution for the aforementioned coded caching problem by designing two arrays: a caching array and a delivery array. Further, we present two classes of caching and delivery arrays and obtain corresponding multi-access coded caching schemes. The first scheme achieves a normalized delivery time (NDT) $\frac{K_1K_2(1-r^2\frac{M}{N})}{L+K_1K_2\frac{M}{N}}$. The second scheme achieves an NDT $\frac{K_1K_2(1-r^2\frac{M}{N})}{L+K_1K_2r^2\frac{M}{N}}$ when $M/N=1/K_1K_2$ and $L=K_1K_2-r^2$, which is optimal under uncoded placement and one-shot delivery.
\end{abstract}

\section{Introduction}
With the advent of smart devices and content streaming applications, network traffic has grown unprecedentedly over the last decade. To combat the traffic congestion experienced during peak hours,
caching is considered to be a promising technique where caches distributed across the network are used to prefetch contents during off-peak hours. When users request a file during peak hours, a part of the contents gets served from the caches. Therefore, the server needs to transmit only the remaining portion, thereby reducing the load over the shared link and the delivery time required to serve the users. In conventional caching techniques, the file contents are sent in uncoded form, and each transmission benefits only a single user. In \cite{MaN}, the authors showed that multicasting opportunities could be created by employing coding in the delivery phase. The network model considered in \cite{MaN} is called a dedicated cache network where there is a server with a library of files and a set of users, each endowed with its own cache, connected through an error-free shared link. The coded caching scheme proposed in \cite{MaN} was shown to be optimal under uncoded placement when the number of files is not less than the number of users \cite{YMA}, \cite{WTP}. Later, the coded caching technique was studied extensively in various settings \cite{HKD, ReK, YCT, STS}.

In this work, we consider the two-dimensional (2D) multi-access coded caching (MACC) model introduced in \cite{ZWCC}. The 2D MACC model is a generalization of the well-studied one-dimensional multi-access coded caching network, first introduced in \cite{HKD}. The one-dimensional MACC network consists of an identical number of users and caches, and each user accesses a set of consecutive caches in a cyclic wrap-around manner. The study of cyclic wrap-around MACC networks was motivated by the circular Wyner model for interference networks \cite{WTS}. Several coded caching schemes and lower bounds characterizing the performance were derived for the one-dimensional MACC network \cite{ReK,SPE,CLWZC,SaR,NaR}. 
The study on 2D MACC network was inspired by the cellular networks architecture in \cite{KHY}. The 2D MACC network defined in \cite{ZWCC} consists of $K_1\times K_2$ cache nodes and $K_1\times K_2$ users, where the caches are placed in a rectangular grid of $K_1$ rows and $K_2$ columns. The users are placed regularly in a square grid such that a user can access $r\times r$ neighbouring caches in a cyclic wrap-around fashion. In \cite{ZWCC}, the authors proposed achievable schemes with rates $\frac{K_1K_2(1-r^2\frac{M}{N})}{K_1K_2\frac{M}{N}}$ and $\frac{K_1K_2(1-r^2\frac{M}{N})}{1+K_1K_2\frac{M}{N}}$ for different parameter settings. The work in \cite{CXZW} also considered a variant of the 2D MACC scheme. However, we consider the 2D MACC network proposed in \cite{ZWCC} along with the possibility of multiple transmit antennas at the server. This setting has not been studied before in the coded caching literature. The multi-antenna setting explored in other coded caching networks \cite{PNR,YWCQC,STS,LaE,NPR,PeR} suggests that using multiple antennas at the server can provide notable performance gains compared to the single antenna variants of the problem. 
 In this regard, we study the 2D MACC network with multiple transmit antennas, and propose a few achievability results. 

\subsection{Contributions}
In this work, we introduce the problem of multi-antenna coded caching in a 2D multi-access network with cycle wrap-around, and make the following technical contributions:
\begin{itemize}
	\item We propose a solution for the 2D multi-antenna MACC problem by constructing two arrays, namely a caching array and a delivery array (Section \ref{arrays}, Section \ref{MA2DMACC}). As the names indicate, the placement phase of the proposed coded caching scheme is determined by the caching array and the delivery phase is according the delivery array. 
	The idea of caching and delivery arrays follows from the extended placement delivery arrays (EPDA) proposed in \cite{NPR}. 
	\item We propose two constructions of caching and delivery arrays. First, we show that caching and delivery arrays can be obtained by manipulating EPDAs if $r|K_1$ (\textit{$r$ divides $K_1$)} and $r|K_2$. From the obtained caching and delivery arrays, we derive an MACC scheme with a normalized delivery time (NDT) $T_n =\frac{K_1K_2(1-r^2\frac{M}{N})}{L+K_1K_2\frac{M}{N}}$ (Lemma \ref{lem:construction}, Remark \ref{rem:const}). 
	\item The optimality of the single-antenna 2D MACC scheme in \cite{ZWCC} itself is unknown. But interestingly, we present an optimal 2D MACC scheme in the multi-antenna setting. The proposed scheme is applicable for the case when $M/N=1/K_1K_2$ and $L=K_1K_2-r^2$, and we present a construction of caching and delivery arrays for the same.
	The resulting MACC scheme has an NDT $T_n=\frac{K_1K_2(1-r^2\frac{M}{N})}{L+K_1K_2r^2\frac{M}{N}}$, which is optimal under uncoded placement and one-shot linear delivery (Corollary \ref{cor1}). Further, with the help of an example, we show that this construction is applicable to a general setting where a) $t|K_1$ and $r\leq K_1/t$ or $t|K_2$ and $r\leq K_2/t$, and b) $L=K_1K_2-r^2t$, $t=K_1K_2M/N$ is an integer (Remark \ref{rem2}). 
\end{itemize}


\subsection{Notations}
For a positive integer $m$, the set $ \left\{1,2,\hdots,m\right\}$ is denoted as $[m]$. For two positive integers $a,b$ such that $a\leq b$, $[a:b] = \{a,a+1,\hdots,b\}$. For integers $a,b\leq K$ 
\begin{equation*}
[a:b]_K =
\begin{cases}
\left\{a,a+1,\hdots,b\right\} & \text{if } a\leq b.\\
\left\{a,a+1,\hdots,K,1,\hdots,b\right\} & \text{if } a>b.
\end{cases}   
\end{equation*}
For any two integers $j$ and $m$, $\langle j \rangle_m $ denotes $j\text{ }(mod\text{ }m)$ if $j$ $(mod$ $m) \neq0$, and $\langle j \rangle_m=m $ if $j$ $(mod$ $m) =0$.
For two positive integers $p$ and $q$, $q|p$ means that $q$ divides $p$ (i.e., $p=zq$ for some integer $z$). For two vectors $\mathbf{u}$ and $\mathbf{v}$, $\mathbf{u}\perp \mathbf{v}$ means that $\mathbf{v}^T\mathbf{u}=0$, and $\mathbf{u}\not\perp \mathbf{v}$ means that $\mathbf{v}^T\mathbf{u}\neq0$. Finally, the symbol $\mathbb{C}$ represents the field of complex numbers. 

\section{System Model}
\subsection{Problem Setting}
\label{subsec:netw}
\begin{figure}[t]
	\begin{center}
		\captionsetup{justification = centering}
		\includegraphics[width = 0.95\columnwidth]{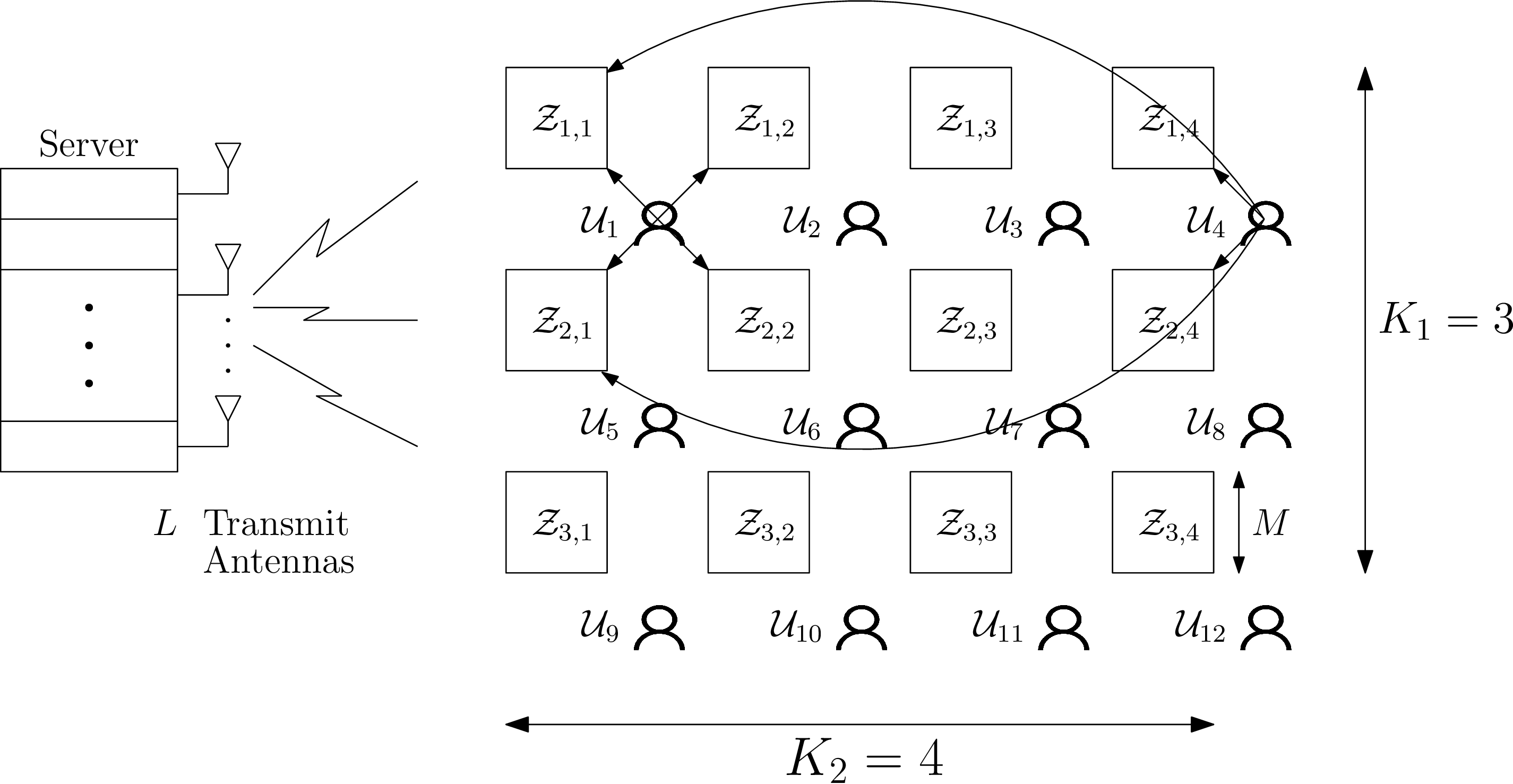}
		\caption{$(K_1,K_2,r,L,M,N)$ MACC system}
		\label{MACC}
	\end{center}
\end{figure}
The system model consists of a central server with $N$ files, $W_{[1:N]}\triangleq \{W_n:n\in [N]\}$, each of size $B$ bits. We consider a multiple-input single-output (MISO) broadcast channel in which a server with $L$ transmit antennas communicates to a set of users, each with a single receive antenna. There are $K_1\times K_2$ cache nodes and as many users in the network, where $K_1$ and $K_2$ are positive integers. The cache nodes, each having a capacity of $MB$ bits, are placed in a rectangular grid consisting of $K_1$ rows and $K_2$ columns. The cache nodes are denoted as $\mathcal{Z}_{(k_1,k_2)}$, where $k_1\in [K_1]$ and $k_2\in [K_2]$. Without loss of generality, we assume that $K_2\geq K_1$. The $K_1K_2$ users, $\{\mathcal{U}_k:k\in [K_1 K_2]\}$, are placed regularly on the square grid such that a user accesses $r\times r$ neighbouring caches in a cyclic wrap-around fashion (we assume that $r<K_1$). User $\mathcal{U}_k$, where $k=(i-1)K_2+j$, $i\in [K_1], j\in [K_2]$, can access all the cache nodes $\mathcal{Z}_{(k_1,k_2)}$ such that $\max(\langle k_1-i\rangle_{K_1},\langle k_2-j \rangle_{K_2})< r$. In other words, user $\mathcal{U}_k$ can access the caches $\mathcal{Z}_{(i+u,j+v)}$ for every $u,v\in [r]$. Further, we assume that the number of files with the server is not less than the number of users. i.e., $N\geq K_1K_2$. We refer to the aforementioned setting as the $(K_1,K_2,r,L,M,N)$ multi-access coded caching (MACC) network. A $(K_1=3,K_2=4,r=2,L,M,N)$ MACC network is shown in Figure \ref{MACC}.   

The $(K_1,K_2,r,L,M,N)$ MACC scheme operates in two phases: a placement phase and a delivery phase. In the placement phase, the server fills the caches with file contents, adhering to the capacity of the caches. The cache placement is done without knowing the demands of the users. Further, the placement phase can be coded or uncoded. In this work, we design schemes with uncoded placement, where the files are divided into subfiles and stored in the caches without making any coded combinations of the files or subfiles. The content stored in $\mathcal{Z}_{(k_1,k_2)}$ is denoted as $Z_{(k_1,k_2)}$. The delivery phase starts after the users' demands are known to the server. Let user $\mathcal{U}_k$ demands the file $W_{d_k}$ from the server, where $k=(i-1)K_2+j$, $i\in [K_1], j\in [K_2]$. The demand of all the users are denoted by $\mathbf{d}=(d_1,d_2,\ldots,d_{K_1K_2})$. After knowing the demand vector $\mathbf{d}$, the server broadcasts the message $\mathbf{X} = (\mathbf{x}(\tau))_{\tau=1}^{T(\mathbf{d})}$ to the users over the channel, where $T(\mathbf{d})$ is the duration or block-length of the channel code, and  $\mathbf{x}(\tau)\in \mathbb{C}^{L \times 1}$ is the transmitted vector at time $\tau\in [T(\mathbf{d})]$ subject to the power constraint $\mathbb{E}(||\mathbf{x}(\tau)||^2)\leq P$. The message $\mathbf{X}$ is a function of $\mathbf{d}$, the cache contents, and the channel coefficients between the transmit antennas and the users. 
At time $\tau$,  the received message at user $\mathcal{U}_k$ is given by
$y_k(\tau) = \mathbf{h}_k^\mathrm{T}\mathbf{x}(\tau)+w_k(\tau)$,
where $\mathbf{h}_k\in\mathbb{C}^{L \times 1}$ denotes the channel
between the server and user $\mathcal{U}_k$, and $w_k(\tau)\sim \mathbb{C}\mathcal{N} (0,1)$ (circularly symmetric gaussian with unit variance) is the
additive noise observed at user $\mathcal{U}_k$ at time $\tau$. 
The entries of the channel matrix $\mathbf{H} = [\mathbf{h}_1,\mathbf{h}_2,\dots,\mathbf{h}_{K_1K_2}]$ are independent and identically distributed, and remain invariant within each codeword transmission. Also, we assume that the server and the users have perfect channel state information (CSI). User $\mathcal{U}_k$ recovers an estimate $\widetilde{W}_{d_k}$ of the demanded file $W_{d_k}$ using $(y_k(\tau))_{\tau=1}^{T(\mathbf{d})}$ and its accessible cache contents. For a given cache placement, we define the worst-case probability of error as $$
P_e  =\max\limits_{\mathbf{d}\in [N]^{K_1K_2}} \max\limits_{k\in [K_1K_2]}\hspace{0.1cm} \mathbb{P}(\widetilde{W}_{d_k}\neq {W}_{d_k}).$$
A coding scheme is said to be feasible if the worst-case probability of error $P_e\rightarrow 0$ as the file size $B\rightarrow \infty$. We use the metric normalized delivery time (defined below) to characterize the performance of the multi-antenna MACC scheme.
\begin{defn}[Normalized delivery time (NDT) \cite{STS2}]
	\label{NDT}
	For a feasible coding scheme, an NDT $T_n$ is said to be achievable if $
	T_n = \lim\limits_{P\rightarrow \infty}\lim\limits_{B\rightarrow \infty} \frac{T}{B/\log P}$,
	where $T =\max\limits_{\mathbf{d}\in [N]^{K_1K_2}} T(\mathbf{d})$ and $B/\log P$ is the time required to transmit $B$ bits to a single user at high SNR with a transmission rate $\log P$ . The optimal NDT is defined as $$
	T_n^* = \inf\{T_n:T_n \text{ is achievable}\}.$$
\end{defn}
The NDT is the worst-case delivery time required to serve all the users' demands normalized with respect to the time required to serve a single file to a single user under cacheless setting. Note that the NDT is identical to the rate measure used for the single-stream error-free setting in \cite{ZWCC}, since it accounts for the file size and the high SNR link capability. 

Now, we revisit the multi-antenna coded caching problem in the dedicated cache network to give a bound on the performance of the multi-antenna scheme in the 2D MACC setting. Consider a dedicated cache network with $K$ users, each having a cache of size $M$ bits and a single receive antenna, connected to a server having $L$ transmit antennas and a library of $N$ files. The optimal NDT for the above network, under uncoded placement and one-shot delivery is $\frac{K(1-M/N)}{L+KM/N}$  \cite{LaE1}. Any achievable NDT of the 2D MACC scheme is also achievable in a dedicated cache network with $L$ transmit antennas and $K_1K_2$ users, each having a cache size of $r^2M$ bits. Hence, under uncoded placement and one-shot delivery, the optimal NDT of the 2D MACC scheme is lower bounded by $T^*_n(M)\geq \frac{K_1K_2(1-r^2M/N)}{L+K_1K_2r^2M/N}$.

 \subsection{Preliminaries}
 \label{prelims}
 In this section, we briefly review the extended placement delivery array (EPDA) in \cite{NPR}, \cite{YWCQC}. Corresponding to an EPDA, there exists a multi-antenna coded caching scheme for the dedicated cache network. The construction of our caching and delivery arrays  makes use of EPDAs, which are defined as follows:
 \begin{defn}[Extended Placement Delivery Array \cite{NPR}]
 	\label{defn:ata}
 	Let $K,L (\leq K), F,Z, S$ be positive integers. An array $\mathbf{A}=[a_{j,k}]$, $j\in [F]$, $k\in [K]$ consisting of the symbol $\star$ and positive integers in $[S]$ is called a $(K,L,F,Z,S)$ extended placement delivery array (EPDA) if it satisfies the following conditions:\\
 	C1. The symbol $\star$ appears $Z$ times in each column.\\
 	C2. Every integer in the set $[S]$ occurs at least once in $\mathbf{A}$.\\ 
 	C3. No integer appears more than once in any column.\\
 	C4. Consider the sub-array $\mathbf{A}^{(s)}$ of $\mathbf{A}$ obtained by deleting all the rows and columns of $\mathbf{A}$ that do not contain the integer $s$. Then for any $s\in [S]$, no row of $\mathbf{A}^{(s)}$ contains more than $L$ integers.
 \end{defn}
\begin{thm}[\cite{NPR}]
	\label{thm:ata}
	Corresponding to any $(K,L,F,Z,S)$ EPDA, there exists a $(K,L,M,N)$ multi-antenna coded caching scheme for the dedicated cache network with ${M}/{N}={Z}/{F}$. Furthermore, the server can meet any user demand $\mathbf{d}$ with an NDT $T_n = {S}/{F}$.
\end{thm}

 \section{Multi-antenna Coded Caching Schemes for 2D MACC Networks}
 \label{MainResults}
 In this section, we present two multi-antenna coded caching schemes for the network model described in Section~\ref{subsec:netw}. The schemes are derived from two arrays, namely a caching array and a delivery array consisting of some special characters and integers. The caching and delivery arrays are defined in the next subsection, and following it we show that there exists a multi-antenna MACC scheme corresponding to every caching array-delivery array pair.
 
 \subsection{Caching and Delivery Arrays}
 \label{arrays}
 \subsubsection{Caching Array}
 Let $K_1,K_2(\geq K_1),F$, and $Z$ be positive integers. An array $\mathbf{C}=[c_{f,(k_1,k_2)}]$, $f\in [F]$, $k_1\in [K_1]$ and $k_2\in [K_2]$ consisting of a special symbol `$\star$' and ` ' (null) is said to be a $(K_1,K_2,F,Z)$ caching array if the symbol $\star$ appears $Z$ times in each column.
 
 \subsubsection{Delivery Array}

 	Let $\mathbf{C}$ be a $(K_1,K_2,F,Z)$ caching array, and let $r,L$, and $S$ be positive integers. An array $\mathbf{B}=[b_{f,(k_1,k_2)}]$, $f\in [F]$, $k_1\in [K_1]$, and $k_2\in [K_2]$, consisting of the symbol $\star$ and positive integers in $[S]$ is called a $(\mathbf{C}(K_1,K_2,F,Z),r,L,S)$ delivery array $\mathbf{B}$ (or a $(\mathbf{C},r,L,S)$ delivery array $\mathbf{B}$) if it satisfies the following conditions:\\  
 	D1: The symbol $\star$ appears in $\mathbf{B}$ according to the $\star$'s in $\mathbf{C}$ as
 	\begin{align*}
 		b_{f,(k_1,k_2)} = \star;&\text{ if } c_{f,(k_1',k_2')}=\star \text{ for some $(k_1',k_2')$ such that }\\& \text{$\max(\langle k_1'-k_1 \rangle_{K_1},\langle k_2'-k_2 \rangle_{K_2})< r$.}
 	\end{align*}
 	D2. Every integer in the set $[S]$ occurs at least once in $\mathbf{B}$.\\ 
 	D3. No integer appears more than once in any column.\\
 	D4. Consider the sub-array $\mathbf{B}^{(s)}$ of $\mathbf{B}$ obtained by deleting all the rows and columns of $\mathbf{B}$ that do not contain the integer $s$. Then, for any $s\in [S]$, no row of $\mathbf{B}^{(s)}$ contains more than $L$ integers.
 	
 	We now give an example of caching array-delivery array pair.
 \begin{exmp}
 	\label{example1}
 	A $(K_1 =3,K_2=3,F=9,Z=1)$ caching array $\mathbf{C}$ is given below:
  	\begin{equation}  
 	\label{exam1:cachingarray}
 	\mathbf{C} =
 	\begin{bmatrix}
 	\star & & &  \\
 	&\star & &  \\
 	& & \ddots &  \\
 	& &  &  \star
 	\end{bmatrix}.
 	\end{equation}
 	The caching array $\mathbf{C}$ is of size $9 \times 9$. The columns of $\mathbf{C}$ are indexed as $(1,1),(1,2),$$(1,3),(2,1),(2,2),$ $(2,3),(3,1),(3,2),(3,3)$ from left to right. Note that each column of $\mathbf{C}$ contains exactly one $\star$ as $Z=1$. Next, we present a $(\mathbf{C},2,5,5)$ delivery array $\mathbf{B}$ in \eqref{exam2:deliveryarray}.
 	 \begin{equation}  
 	 \label{exam2:deliveryarray}
 	 \mathbf{B} =
 	 \begin{bmatrix}
\star&1    &\star&1    &1    &1    &\star&1    &\star \\
\star&\star&1    &2    &2    &2    &\star&\star&1     \\
1    &\star&\star&3    &3    &3    &1    &\star&\star \\
\star&2    &\star&\star&4    &\star&2    &2    &2     \\
\star&\star&2    &\star&\star&4    &3    &3    &3     \\
2    &\star&\star&4    &\star&\star&4    &4    &4     \\
3    &3    &3    &\star&5    &\star&\star&5    &\star \\
4    &4    &4    &\star&\star&5    &\star&\star&5     \\
5    &5    &5    &5    &\star&\star&5    &\star&\star
 	 \end{bmatrix}.
 	 \end{equation}
 	 The delivery array $\mathbf{B}$ is of same size as $\mathbf{C}$, and the columns of $\mathbf{B}$ are indexed as: $(1,1),(1,2),(1,3),(2,1),(2,2),(2,3),(3,1),(3,2),(3,3)$ from left to right, as is done in $\mathbf{C}$. The $\star$'s in any column $(k_1,k_2)$ of $\mathbf{B}$, $k_1 \in [3], k_2 \in [3]$, is determined by the $\star$'s in the following columns of $\mathbf{C}$:  $(k_1,k_2)$,$(k_1,\langle k_2+1 \rangle_3)$,$(\langle k_1+1 \rangle_3,k_2)$ and $(\langle k_1+1\rangle_3,\langle k_2+1\rangle_3)$ . Notice that $\star$'s in a column of $\mathbf{B}$ is determined by $\star$'s in $r^2=4$ columns of $\mathbf{C}$. The delivery array consists of integers from the set $[5]$, and each integer appears 9 times in the array. Also, no integer appears more than once in any column. Now, consider sub-array $\mathbf{B}^{(1)}$ corresponding to integer 1.
 	\begin{equation*}
  	 \mathbf{B}^{(1)}=
 	 \begin{bmatrix}
 	 		\star&1    &\star&1    &1    &1    &\star&1    &\star \\
 	 		\star&\star&1    &2    &2    &2    &\star&\star&1     \\
 	 		1    &\star&\star&3    &3    &3    &1    &\star&\star  	 	
 	 \end{bmatrix}.
  \end{equation*}
Note that no row of $\mathbf{B}^{(1)}$ contains more than $L=5$ integers. This is true for every sub-array $\mathbf{B}^{(s)}$, $s\in [5]$. 	 	
 \end{exmp}
 
 In the following subsection, we show that corresponding to any caching array-delivery array pair, we can obtain a multi-antenna 2D MACC scheme. 
 \subsection{2D MACC Schemes with Multiple Transmit Antennas}
 \label{MA2DMACC}
 Let $\mathbf{C}$ be a $(K_1,K_2,F,Z)$ caching array and $\mathbf{B}$ be a corresponding $(\mathbf{C},r,L,S)$ delivery array. From this caching and delivery arrays, by employing the following placement and delivery procedures, we obtain a $(K_1,K_2,r,L,M,N)$ 2D MACC scheme with a normalized cache size $M/N=Z/F$ and an NDT $T_n=S/F$. Each column of the caching array $\mathbf{C}$ represents a cache node, and each column of the delivery array $\mathbf{B}$ represents a user. The placement and delivery procedures are as follows:
 
 a) Placement phase: The cache placement is done according to the $(K_1,K_2,F,Z)$ caching array $\mathbf{C}$. In the placement phase, each file is divided into $F$ subfiles: $W_n = \{W_{n,f}: f\in [F]\}, \forall n\in [N]$. Then, the server fills cache $\mathcal{Z}_{(k_1,k_2)}$ as follows:
 \begin{equation*}
 Z_{(k_1,k_2)} = \{W_{n,f}, \forall n\in [N]: c_{f,(k_1,k_2)}=\star, f\in [F]\}.
 \end{equation*}
 The rows of $\mathbf{C}$ represent the subpacketization level required by the scheme. By definition, each column of a caching array contains $Z$ $\star$'s. Therefore, a cache stores $Z$ subfiles of every file. Thus, we have $M/N=Z/F$. Consider a user $\mathcal{U}_k$, where $k \in [K_1,K_2]$ and $k=(i-1)K_2+j$, $i\in [K_1], j\in [K_2]$. User $\mathcal{U}_k$ has access to the contents of the caches indexed by $(k_1,k_2)$ such that $\max(\langle k_1-i\rangle_{K_1},\langle k_2-j \rangle_{K_2})< r$. The $\star$'s in a column of the delivery array $\mathbf{B}$ represent the subfiles known to the corresponding  user.\\
 b) Delivery phase: The transmissions in the delivery phase is determined by the $(\mathbf{C},r,L,S)$ delivery array $\mathbf{B}$. Condition D1 ensures that the $\star$'s in the column indexed with ordered pair $(i,j)$ of the delivery array $\mathbf{B}$ locate the subfiles accessible to user $\mathcal{U}_k$, where $k=(i-1)K_2+j$. Further, if $b_{f,(i,j)} = s$ for some $s\in [S]$, then user $\mathcal{U}_k$ requires the subfile $W_{d_k,f}$ from the delivery phase. 
 
 Assume that integer $s$ appears $\eta_s$ times in $\mathbf{B}$, where $s\in [S]$. Let $b_{f_1,(i_1,j_1)}=b_{f_2,(i_2,j_2)}=\dots=b_{f_{\eta_s},(i_{\eta_s},j_{\eta_s})}=s$. Also, define $k^{(\alpha)}\triangleq(i_{\alpha}-1)K_2+j_{\alpha}$ for every $\alpha \in [\eta_s]$. Notice that, $\mathcal{U}_{k^{(\alpha)}}$ is the user corresponding to column $(i_\alpha,j_\alpha)$ of $\mathbf{B}$. First, the server creates a precoding matrix $\mathbf{V}^{(s)}=[\mathbf{v}_1^{(s)},\mathbf{v}_2^{(s)},\dots,\mathbf{v}_{\eta_s}^{(s)}]$  of size $L\times \eta_s$ as follows: for every $\alpha \in [\eta_s]$, define a set $\mathcal{P}_\alpha=\{k^{(\beta)}:d_{f_\alpha,(i_\beta,j_\beta)}\neq \star, \beta\in [\eta_s]\backslash\{\alpha\}\}$. Then the $\alpha^{th}$ column of $\mathbf{V}^{(s)}$ is such that $\mathbf{v}_\alpha^{(s)}\not\perp \mathbf{h}_{k^{(\alpha)}}$ and $\mathbf{v}_\alpha^{(s)}\perp \mathbf{h}_{k^{(\gamma)}}$ for every $\gamma \in \mathcal{P}_\alpha$. The server broadcasts $\mathbf{V}^{(s)}(W_{d_{k^{(1)}},f_1},W_{d_{k^{(2)}},f_2},\dots,W_{d_{k^{(\eta_s)}},f_{\eta_s}})^T$ over the channel. Note that, condition D4 ensures that $|\mathcal{P}_\alpha|\leq L-1$. Thus, for every $\alpha\in [\eta_s]$, it is possible to construct the precoding vector $\mathbf{v}_\alpha^{(s)}$ with probability 1. 
 By employing the above placement and delivery procedures, we have the following theorem.
 
 \begin{thm}
 	\label{thm:array}
 	Corresponding to a $(K_1,K_2,F,Z)$ caching array $\mathbf{C}$ and a $(\mathbf{C},r,L,S)$ delivery array $\mathbf{B}$, there exists a $(K_1,K_2,r,L,M,N)$ MACC scheme with a normalized cache memory $M/N=Z/F$. Further, the server can meet any user demand with an NDT $T_n=S/F$.
 \end{thm}
 \begin{IEEEproof}
 	In the placement phase, each file is divided into $F$ subfiles, and $Z$ of them are stored in the caches. Therefore, the normalized cache size is $M/N=Z/F$. Further, in the delivery phase, there is a transmission corresponding to every integer in $[S]$. Each transmission incurs a normalized delay of $1/F$, as each transmission has a size $(\frac{1}{F})^{th}$ of a file size. Thus, the NDT is $T_n=S/F$. Now, it remains to show that every user can recover its demanded file.
 	
 	Consider user $\mathcal{U}_k$ that wants the subfile $W_{d_k,f}$, where $k =(i-1)K_2+j$ for some $i\in [K_1]$ and $j\in [K_2]$. Let $b_{f,(i,j)}=s$, for some $s\in [S]$. If $b_{f,(i,j)}= \star$, then that subfile would have been available to $\mathcal{U}_k$ from the placement phase itself. Now, we show that user $\mathcal{U}_k$ can recover $W_{d_k,f}$ from the transmission corresponding to integer $s$. In the sub-array $\mathbf{B}^{(s)}$, let $b_{f,(i,j)}=b_{f_2,(i_2,j_2)}=\dots=b_{f_{\eta_s},(i_{\eta_s},j_{\eta_s})}=s$. From the server transmission $\mathbf{x}{(s)} =\mathbf{V}^{(s)}(W_{d_{k},f},W_{d_{k^{(2)}},f_2},\dots,W_{d_{k^{(\eta_s)}},f_{\eta_s}})^T$, user $\mathcal{U}_k$ receives:
 	\begingroup
 	\allowdisplaybreaks
 	\begin{align*}
 	y_k(s) &= \mathbf{h}_k^{\mathrm{T}}\mathbf{V}^{(s)}(W_{d_{k},f},W_{d_{k^{(2)}},f_2},\dots,W_{d_{k^{(\eta_s)}},f_{\eta_s}})^{\mathrm{T}}
 	\\&=(\mathbf{h}_k^{\mathrm{T}} \mathbf{v}_1^{(s)}) W_{d_k,f}+\underbrace{\sum\limits_{\substack{\alpha=2 \\ b_{f_\alpha,(i,j) }=\star}}^{\eta_s} \left(\mathbf{h}_k^{\mathrm{T}} \mathbf{v}_\alpha^{(s)}\right) W_{d_{k^{(\alpha)}},f_\alpha}.}_{\text{$\mathcal{U}_k$ can compute using accessible cache contents}}
 	\end{align*}
 	\endgroup	 
 	Notice that, we neglect the additive noise $w_k(s)$ in the further analysis due to the high SNR assumption. The design of precoding vectors ensures that $\mathbf{h}_k^{\mathrm{T}} \mathbf{v}_\alpha^{(s)}=0$ for all $\alpha\in [2:\eta_s]$ such that $b_{f_\alpha,(i,j) }\neq\star$. In fact, the precoding vectors are designed such that the subfiles that are not known and are not required for user $\mathcal{U}_k$ will be nulled out in ${y}_k(s)$, $\forall s \in [s]$. Thus, $\mathcal{U}_k$ gets the required subfile $W_{d_k,f}$. Note that $k$ and $f$ are arbitrary. Therefore, all the users can recover their demanded files. This completes the proof of Theorem \ref{thm:array}.
  \end{IEEEproof}

\begin{exmp}
	\label{exam:scheme}
	Consider the $(3,3,9,1)$ caching array $\mathbf{C}$ in \eqref{exam1:cachingarray} and the $(\mathbf{C},2,5,5)$ delivery array $\mathbf{B}$ in \eqref{exam2:deliveryarray}. From the $(\mathbf{C},\mathbf{B})$ pair, we obtain a $(K_1=3,K_2=3,r=2,L=5,M=1,N=9)$ MACC scheme. The server has $9$ files $\{W_1,W_2,\dots,W_9\}$. The server communicates to the users with $L=5$ transmit antennas. 
	The network consists of $9$ caches and $9$ users arranged in a $3 \times 3$ rectangular grid, where each user accesses 4 neighbouring caches in a cyclic wrap-around manner.  The corresponding 2D MACC network is shown in Figure~\ref{example2}. The set of caches accessible to each user are listed below:
	\begin{align*}
	\mathcal{U}_1 &\textrm{ accesses } \{\mathcal{Z}_{1,1},\mathcal{Z}_{1,2},\mathcal{Z}_{2,1},\mathcal{Z}_{2,2}\},\\
	\mathcal{U}_2 &\textrm{ accesses } \{\mathcal{Z}_{1,2},\mathcal{Z}_{1,3},\mathcal{Z}_{2,2},\mathcal{Z}_{2,3}\},\\
	\mathcal{U}_3 &\textrm{ accesses } \{\mathcal{Z}_{1,3},\mathcal{Z}_{1,1},\mathcal{Z}_{2,3},\mathcal{Z}_{2,1}\},\\
	\mathcal{U}_4 &\textrm{ accesses } \{\mathcal{Z}_{2,1},\mathcal{Z}_{2,2},\mathcal{Z}_{3,1},\mathcal{Z}_{3,2}\}, \\
	\mathcal{U}_5 &\textrm{ accesses } \{\mathcal{Z}_{2,2},\mathcal{Z}_{2,3},\mathcal{Z}_{3,2},\mathcal{Z}_{3,3}\}, \\
	\mathcal{U}_6 &\textrm{ accesses } \{\mathcal{Z}_{2,3}, \mathcal{Z}_{2,1},\mathcal{Z}_{3,3},\mathcal{Z}_{3,1}\}, \\
	\mathcal{U}_7 &\textrm{ accesses } \{\mathcal{Z}_{3,1},\mathcal{Z}_{3,2}, \mathcal{Z}_{1,1},\mathcal{Z}_{1,2}\}, \\
	\mathcal{U}_8 &\textrm{ accesses } \{\mathcal{Z}_{3,2}, \mathcal{Z}_{3,3},\mathcal{Z}_{1,2},\mathcal{Z}_{1,3}\},\\
	\mathcal{U}_9 &\textrm{ accesses } \{\mathcal{Z}_{3,3}, \mathcal{Z}_{3,1},\mathcal{Z}_{1,3},\mathcal{Z}_{1,1}\}.
	\end{align*}

	\begin{figure}[h]
		\begin{center}
			\captionsetup{justification = centering}
			\includegraphics[width = \columnwidth]{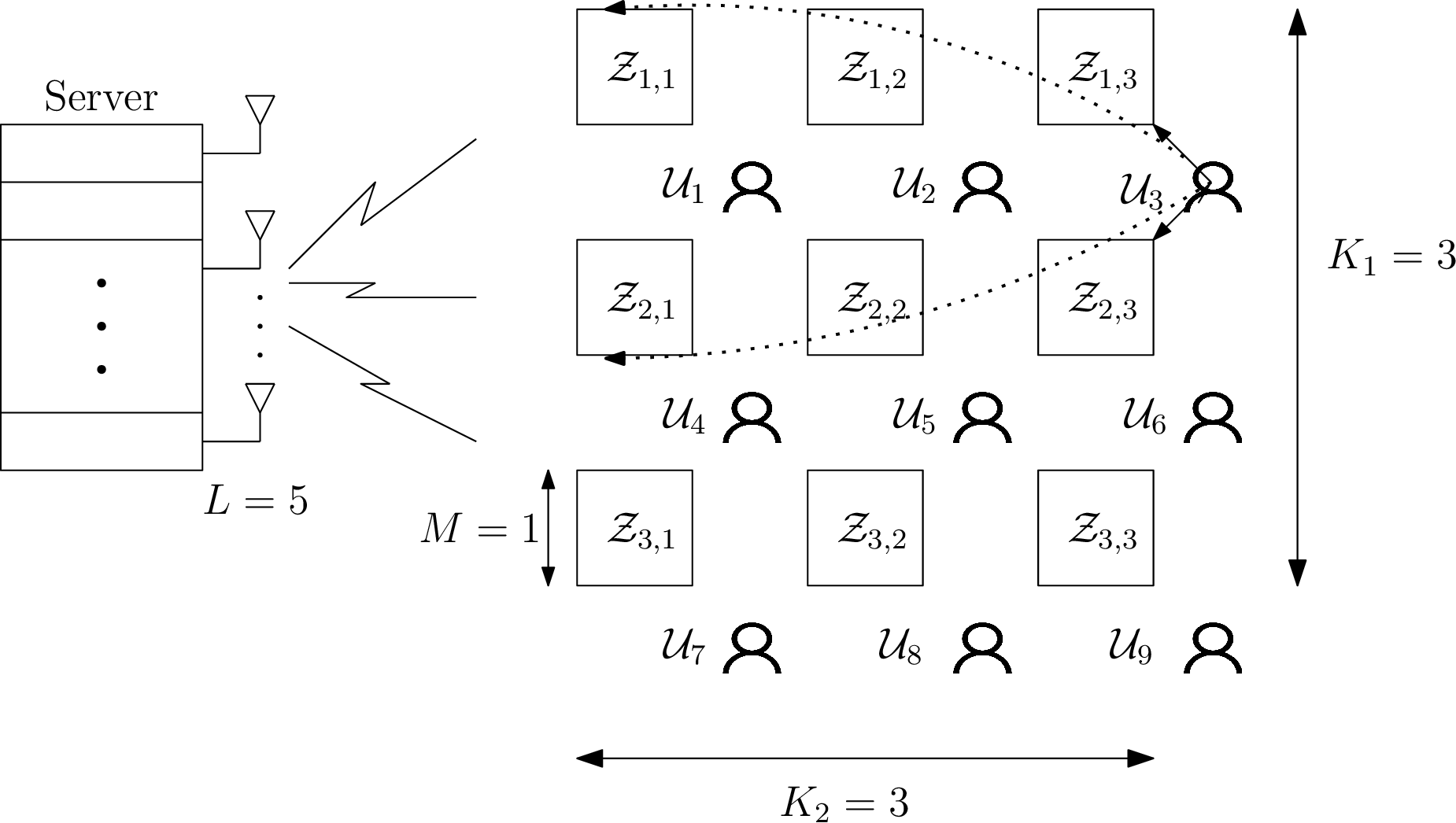}
			\caption{$(3,3,2,5,1,9)$ 2D MACC network}
			\label{example2}
		\end{center}
	\end{figure}

	i) Placement phase: The server divides each file into $F=9$ subfiles, $W_{n}=\{W_{n,f},f\in [9]\} , \forall n\in [9]$. Since $C_{1,(1,1)}=\star$, cache $\mathcal{Z}_{(1,1)}$ is filled as $Z_{(1,1)}=\{W_{n,1},n\in [9]\}$, . Similarly, the other caches are filled as $Z_{(1,2)}=\{W_{n,2},n\in [9]\},Z_{(1,3)}=\{W_{n,3},n\in [9]\},\dots,Z_{(3,3)}=\{W_{n,9},n\in [9]\}$. Each user can access $r^2=4$ caches. For instance, $\mathcal{U}_3$ accesses caches $\mathcal{Z}_{(1,3)},\mathcal{Z}_{(2,3)},\mathcal{Z}_{(1,1)}$, and $\mathcal{Z}_{(2,1)}$. \\
	ii) Delivery phase: Let $\mathbf{d}=(d_1,d_2,\dots,d_9)$ be the demand vector. There are $S=5$ transmissions from the server, each taking a normalized time of $1/F=1/9$. The transmission corresponding to integer 1 is
   \begin{align*}
     \mathbf{x}(1)=\mathbf{V}^{(1)}(W_{d_1,3},W_{d_2,1},W_{d_3,2},&W_{d_4,1},W_{d_5,1},W_{d_6,1},\\&W_{d_7,3},W_{d_8,1},W_{d_9,2})^T,
   \end{align*}
   where $\mathbf{V}^{(1)}=(\mathbf{v}_1^{(1)},\mathbf{v}_2^{(1)},\dots,\mathbf{v}_9^{(1)})\in \mathbb{C}^{5\times 9}$. Let us see how the precoding vector $\mathbf{v}_1^{(1)}$ is constructed. In the first column of $\mathbf{B}^{(1)}$, we have $b^{(1)}_{3,(1,1)}=1$. In the third row of $\mathbf{B}^{(1)}$, except $b^{(1)}_{3,(1,1)}$, the entries $b^{(1)}_{3,(2,1)}$, $b^{(1)}_{3,(2,2)}$, $b^{(1)}_{3,(2,3)}$, and $b^{(1)}_{3,(3,1)}$ are integers. Therefore, $\mathbf{v}_1^{(1)}$ is designed such that $\mathbf{v}_1^{(1)}\perp \mathbf{h}_4$, $\mathbf{v}_1^{(1)}\perp \mathbf{h}_5$, $\mathbf{v}_1^{(1)}\perp \mathbf{h}_6$, and $\mathbf{v}_1^{(1)}\perp \mathbf{h}_7$ (Note that column indices $(2,1)$, $(2,2)$, $(2,3)$, and $(3,1)$ correspond to users $\mathcal{U}_4$, $\mathcal{U}_5$, $\mathcal{U}_6$, and $\mathcal{U}_7$, respectively). Finally, we have $\mathbf{v}_1^{(1)}\not\perp \mathbf{h}_1$. In a similar manner, the other precoding vectors are designed. Therefore, the NDT is obtained as $T_n=5/9$.  
   
   Now, we see how a user recovers the desired subfiles from the associated transmissions. For instance, we consider user $\mathcal{U}_1$. Since  $b_{3,(1,1)}=1$, the user needs $W_{d_1,3}$ from the delivery phase. Therefore, the user gets $W_{d_1,3}$ from the transmission corresponding to integer 1. From that transmission, the user receives $
   y_1(1)= \mathbf{h}_1^{\mathrm{T}}\mathbf{x}(1)= \mathbf{h}_1^{\mathrm{T}} \mathbf{V}^{(1)}(W_{d_1,3},W_{d_2,1},\dots,W_{d_9,2})^{\mathrm{T}}$. By the design of the precoding vector $\mathbf{v}_7^{(1)}$, we have $\mathbf{h}_1^{\mathrm{T}}\mathbf{v}_7^{(1)}=0$. Therefore,
   $y_1(1) = \mathbf{h}_1^{\mathrm{T}}\mathbf{v}_1^{(1)}W_{d_1,3}+\mathbf{h}_1^{\mathrm{T}}\mathbf{v}_2^{(1)}W_{d_2,1}+\mathbf{h}_1^{\mathrm{T}}\mathbf{v}_3^{(1)}W_{d_3,2}+\mathbf{h}_1^{\mathrm{T}}\mathbf{v}_4^{(1)}W_{d_4,1}+\mathbf{h}_1^{\mathrm{T}}\mathbf{v}_5^{(1)}W_{d_5,1}+\mathbf{h}_1^{\mathrm{T}}\mathbf{v}_6^{(1)}W_{d_6,1}+\mathbf{h}_1^{\mathrm{T}}\mathbf{v}_8^{(1)}W_{d_8,1}+\mathbf{h}_1^{\mathrm{T}}\mathbf{v}_9^{(1)}W_{d_9,2}.$
   Since user $\mathcal{U}_1$ has access to $W_{n,1},W_{n,2}, \forall n\in [N]$, it can eliminate the remaining summands and get $\mathbf{h}_1^{\mathrm{T}}\mathbf{v}_1^{(1)}W_{d_1,3}$. Thus,$\mathcal{U}_1$ retrieves the subfile $W_{d_1,3}$ (note that, perfect CSI is available to all the users). Similarly, all other users can also recover their required subfiles. 
\end{exmp}

  Next, we present two constructions of caching and delivery arrays, and describe the 2D MACC schemes resulting from them.
 \subsection{Array Constructions and the resulting 2D MACC schemes}
 First, we illustrate a procedure to obtain caching and delivery arrays from an EPDA. Using that procedure, we obtain caching and delivery arrays from the class of EPDAs constructed in \cite{NPR}, and then derive 2D multi-access coded caching schemes from them. Those resulting MACC schemes have an NDT $ \frac{K_1K_2(1-r^2M/N)}{L+K_1K_2M/N}$. Next, we give an explicit construction of caching and delivery arrays for a specific set of parameters. Interestingly, the resulting MACC scheme has an NDT $ \frac{K_1K_2(1-r^2M/N)}{L+K_1K_2r^2M/N}$, which is optimal under uncoded placement and one-shot delivery.  
 
 \subsubsection{ Caching and delivery arrays from EPDAs}
 \begin{lem}
 	\label{lem:construction}
 	Assume that $r|K_1$ and $r|K_2$. Then from a $(\frac{K_1K_2}{r^2},L,F,Z,S)$ EPDA, it is possible to obtain a $(K_1,K_2,r^2F,Z)$ caching array $\mathbf{C}$ and a $(\mathbf{C},r,L,r^4S)$ delivery array $\mathbf{B}$.
 \end{lem}
 \begin{IEEEproof}
 	  \begin{figure*}[htbp]
 		\begin{equation}
 		\label{darray}
 		\mathbf{B} =
 		\begin{bmatrix}
 		\mathbf{A} & \pi_{r^2+1}(\mathbf{A})+r^2S & \dots&\pi_{r^4-r^2+1}(\mathbf{A})+(r^4-r^2)S\\
 		\pi_2(\mathbf{A})+S & \mathbf{A}+(r^2+1)S &\dots &\pi_{r^4-r^2+2}(\mathbf{A})+(r^4-r^2+1)S\\
 		\vdots & \vdots& \ddots & \vdots\\
 		\pi_{r^2}(\mathbf{A})+(r^2-1)S &\pi_{2r^2}(\mathbf{A})+(2r^2-1)S & \dots&\mathbf{A}+(r^4-1)S
 		\end{bmatrix}
 		\end{equation}
 		\hrule
 	\end{figure*} 
 	We present an explicit construction of $\mathbf{C}$ and $\mathbf{B}$ from a $(\frac{K_1K_2}{r^2},L,F,Z,S)$ EPDA $\mathbf{A}$. Let $\tilde{\mathbf{A}}$ be an array obtained by replacing all the integers in $\mathbf{A}$ with null. We diagonally concatenate the array $\tilde{\mathbf{A}}$ $r^2$ times to obtain $\mathbf{C}$ (the entries in $\mathbf{C}$ are $\star$'s and null). That is,
 	$$
 	\mathbf{C} =
 	\begin{bmatrix}
 	\tilde{\mathbf{A}} & & \\
 	& \ddots & \\
 	& & \tilde{\mathbf{A}}
 	\end{bmatrix}.$$
 	The array $\mathbf{C}$ has $K_1K_2$ columns and $r^2F$ rows. Further, note that $\tilde{\mathbf{A}}$ has $Z$ number of $\star$'s in each column. Therefore, $\mathbf{C}$ also has $Z$ number of $\star$'s in each column. Therefore $\mathbf{C}$ is a $(K_1,K_2,r^2F,Z)$ caching array. Now, we index the columns of $\mathbf{C}$ in a peculiar way. The delivery array depends on the column indexing of the caching array (through condition D1). Consider two positive integers $u,v\in [r]$. A group of $K_1K_2/r^2$ columns are indexed with ordered pairs $((x-1)r+u,(y-1)r+v)$ for every $x\in [K_1/r]$ and $y\in [K_2/r]$. By a group of $K_1K_2/r^2$ columns, we mean the columns corresponding to one $\tilde{\mathbf{A}}$ (out of the $r^2$ possibilities). There are $r^2$ different ways of choosing $(u,v)$ and there $r^2$ such groups of columns. Within a group, the columns are indexed in the lexicographic order.  
 	
 	For an EPDA $\mathbf{A}$ and a positive integer $q$, the array $\mathbf{A}+q$ is obtained by replacing every integer in the array by the sum of the integer and $q$. That is, integer $s$ in $\mathbf{A}$ is replaced by $s+q$ for every $s\in [S]$. Note that the $\star$'s remain as it is in $\mathbf{A}+q$. Also, for every $\ell\in [r^4]$, $\pi_\ell(\mathbf{A})$ represents the EPDA obtained after a column permutation of $\mathbf{A}$ (note that, an EPDA remains an EPDA with the same parameters even after column permutation). Then the delivery array $\mathbf{B}$  resulting from $\mathbf{C} $ has $\star$'s and integers in the form given in \eqref{darray}. The number of integers in $\mathbf{B}$ is $r^4S$. Also, array $\mathbf{B}$ satisfies D3 and D4, since EPDA $\mathbf{A}$ satisfies C3 and C4, respectively. Therefore, $\mathbf{B}$ is a $(\mathbf{C},r,L,r^4S)$ delivery array. This completes the proof of Lemma \ref{lem:construction}. 
  \end{IEEEproof}
We illustrate the above construction through an example.
\begin{exmp}
	Let $K_1=K_2=4, r=2, L=2$, and $M/N=1/8$. Let $\mathbf{A}$ be a $(4,2,4,2,2)$ EPDA. The EPDA $\mathbf{A}$ and an array $\tilde{\mathbf{A}}$ which is obtained by replacing all the integers in $\mathbf{A}$ with null are given below. $$
	\mathbf{A} =
	\begin{bmatrix}
	\star &2 &1 &\star\\
	\star&\star &2 &1\\
	1&\star & \star  &2 \\
	2& 1 & \star& \star
	\end{bmatrix} \textrm{ and }  \tilde{\mathbf{A}} =
	\begin{bmatrix}
	\star & & &\star\\
	\star&\star & &\\
	&\star & \star  & \\
	&  & \star& \star
	\end{bmatrix}.$$
By employing the technique illustrated in the proof of Lemma \ref{lem:construction} on EPDA $\mathbf{A}$, we obtain a $(4,4,16,2)$ caching array $\mathbf{C}$ as in Figure \ref{Cmatrix} and a $(\mathbf{C},2,2,32)$ delivery array $\mathbf{B}$ as in Figure \ref{Dmatrix}.

	\begin{figure}[h]
		\begin{center}
			\captionsetup{justification = centering}
			\includegraphics[width = \columnwidth]{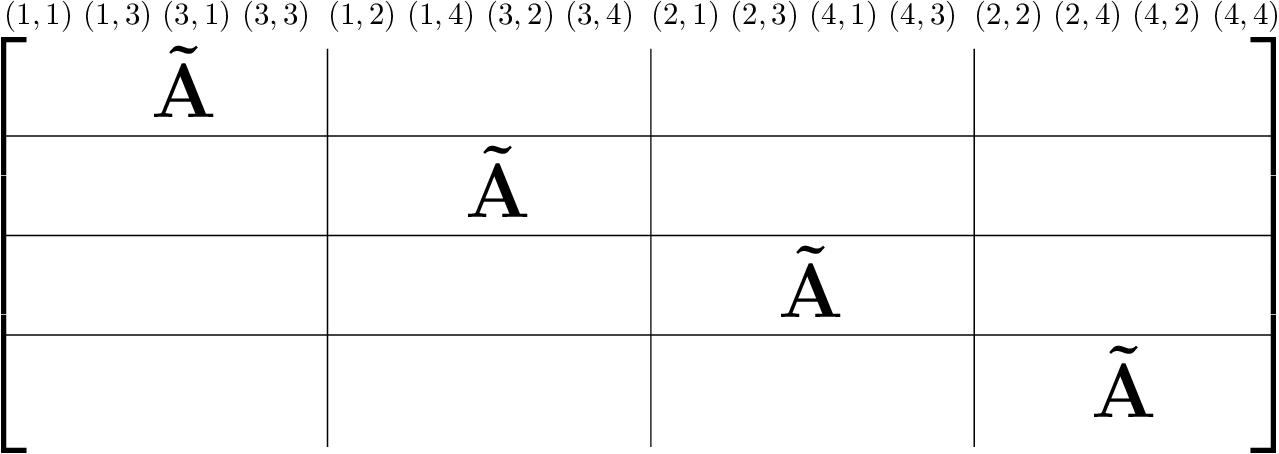}
			\caption{$(4,4,16,2)$ caching array $\mathbf{C}$}
			\label{Cmatrix}
		\end{center}
	\end{figure}

\begin{figure*}[h]
	\begin{center}
		\captionsetup{justification = centering}
		\includegraphics[width = 0.7\textwidth]{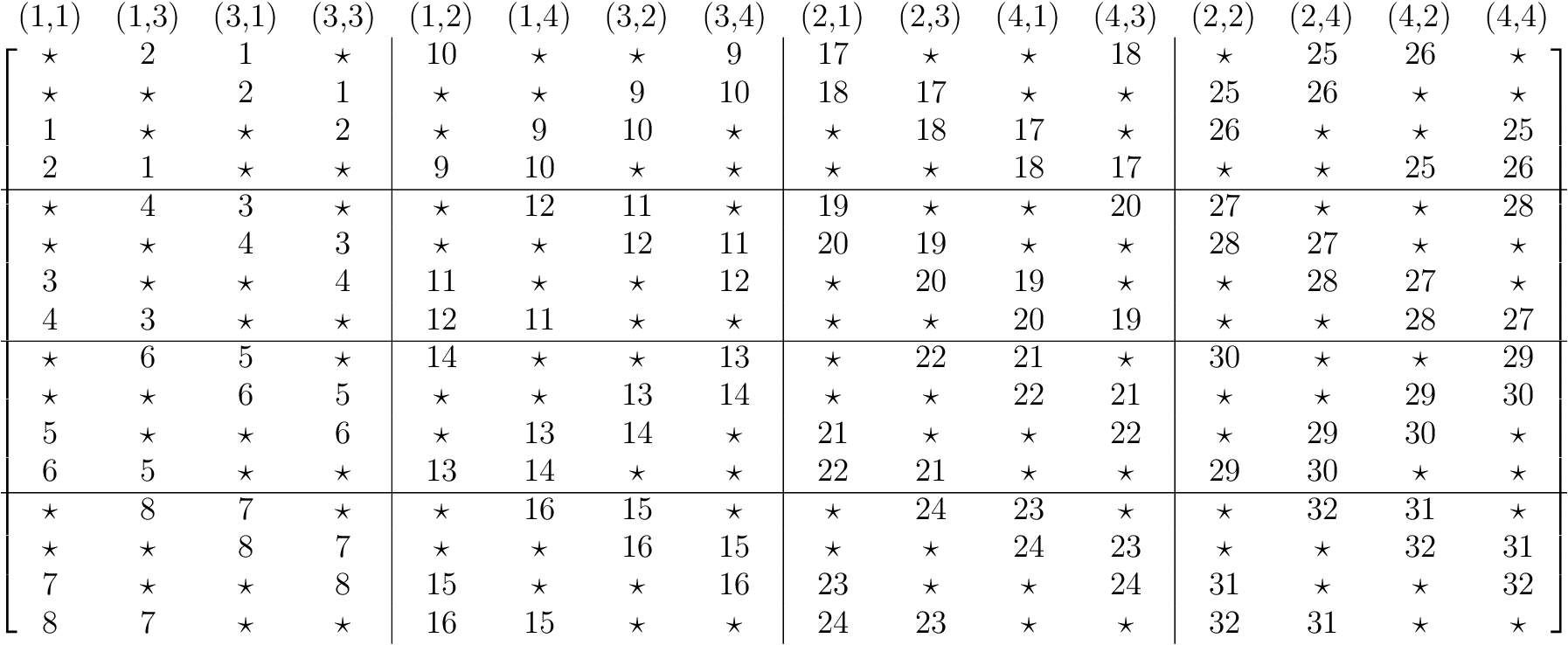}
		\caption{$(\mathbf{C},2,2,32)$ delivery array $\mathbf{B}$}
		\label{Dmatrix}
	\end{center}
\end{figure*}

\end{exmp}

\begin{rem}
	\label{rem:const}
	For positive integers $K,t$, and $L$ such that $t+L\leq K$, an EPDA construction is proposed in \cite{NPR}. Now, let us construct caching and delivery arrays from that class of EPDAs using Lemma \ref{lem:construction}. We consider the case $r|K_1$ and $r|K_2$. Define, $K=K_1K_2/r^2$ and let $M/N$ is such that $t=K_1K_2M/N$ is an integer.
	We have a $(K,L,(t+L)\binom{K}{t+L},t\binom{K-1}{t+L-1},(K-t)\binom{K}{t+L})$ EPDA $\mathbf{A}$ from \cite{NPR}. By applying Lemma \ref{lem:construction}, we get a $(K_1,K_2,r^2(t+L)\binom{K}{t+L},t\binom{K-1}{t+L-1})$ caching array $\mathbf{C}$ and a $(\mathbf{C},r,L,r^4(K-t)\binom{K}{t+L})$ delivery array $\mathbf{B}$. From $\mathbf{C}$ and $\mathbf{B}$, we can obtain a $(K_1,K_2,r,L,M,N)$ MACC scheme with an NDT $\frac{r^2(K-t)}{t+L} =\frac{K_1K_2(1-r^2M/N)}{L+K_1K_2M/N}$.
\end{rem}

 In the MACC schemes obtained from EPDAs, the number of users simultaneously served is limited to $K_1K_2M/N+L$. This happens because of the user grouping. The multicasting is limited to users within a group. Next, we identify an instance of optimality of the $(K_1,K_2,r,L,M,N)$ MACC scheme. In that case, the number of users simultaneously served is $K_1K_2r^2M/N+L$.

 \subsubsection{Construction resulting in an optimal MACC scheme}
 \label{subsec:opt_const}

 The following construction of caching and delivery arrays is valid if $Z/F=1/K_1K_2$ and $r^2+L=K_1K_2$. First, we construct a $(K_1,K_2,K_1K_2,1)$ caching array $\mathbf{C}_1$ as follows: $$\mathbf{C}_1 =
 \begin{bmatrix}
 \star & & &\\
    &\star & &\\
 & & \ddots & \\
 &  & & \star
 \end{bmatrix}.$$
 The columns are indexed in the lexicographic order as $(1,1),(1,2),\dots,(1,K_2),\dots,(K_1,K_2)$. Each row and each column of $\mathbf{C}_1$ contains only a single $\star$, therefore, in the corresponding delivery array $\mathbf{B}_1$, each column contains $r^2$ $\star$'s. The delivery array $\mathbf{B}_1$ contains $S=L$ integers. All integers in the set $[L]$  appears in every column of $\mathbf{B}_1$. This is possible because there are only $r^2=K_1K_2-L$ number of $\star$'s in each column, and the remaining $L$ positions are vacant to fill with integers. We show that $\mathbf{B}_1$ is a delivery array. Since $\mathbf{B}_1$ is obtained from caching array $\mathbf{C}_1$, condition D1 is satisfied. Also, every integer in $[L]$ appears once in every column, and exactly $K_1K_2$ times in the entire array. Therefore, conditions D2 and D3 are also satisfied by $\mathbf{B}_1$. Note that, each row of $\mathbf{B}_1$ contains some $L$ integers. Further, note that each integer in $[L]$ is present in all the $K_1K_2$ columns. Therefore, for every $s\in [S]$, every row of $\mathbf{B}_1^{(s)}$ contains exactly $L$ integers, satisfying D4. Thus, $\mathbf{B}_1$ is a $(\mathbf{C}_1,r,L,S)$ delivery array. 

\begin{corollary}
	\label{cor1}
	For the $(K_1,K_2,r,L=K_1K_2-r^2,M,N)$ MACC scheme with $M/N=1/K_1K_2$, an NDT $\frac{K_1K_2(1-r^2M/N)}{L+K_1K_2r^2M/N}$ is achievable, which is optimal under uncoded placement and one-shot delivery.
\end{corollary}
\begin{IEEEproof}
	Applying Theorem \ref{thm:array} on the above constructed caching array $\mathbf{C}_1$ and delivery array $\mathbf{B}_1$ gives the required $(K_1,K_2,r,L=K_1K_2-r^2,M,N)$ MACC scheme with $M/N=1/K_1K_2$. The NDT achieved is $S/F=L/K_1K_2=K_1K_2(1-r^2M/N)/(L+K_1K_2r^2M/N)$. We have already seen that $T_n^*\geq K_1K_2(1-r^2M/N)/(L+K_1K_2r^2M/N)$. Therefore, the achieved NDT is optimal.
\end{IEEEproof}

 \begin{rem}
 	\label{rem2}
 	Let $K_1,K_2,M$ and $N$ be such that $t \triangleq K_1K_2M/N$ is an integer. The previous construction of caching and delivery arrays is also applicable if a) $t|K_1$ and $r\leq K_1/t$ or $t|K_2$ and $r\leq K_2/t$, b) $r^2t+L=K_1K_2$.
 	 We illustrate this with the help of the following example.
 \end{rem}

\begin{exmp}
	Consider the  $(K_1=3, K_2=4,r=2,L=4,M,N)$ MACC scheme with $M/N=1/6$. Therefore, we have $t=K_1K_2M/N=2$. Also, note that $t|K_2$ and $r\leq K_2/t$. Now, we construct a $(K_1=3,K_2=4,F=K_1K_2/t=6,Z=1)$ caching array $\mathbf{C}_1$ of size $6 \times 12$ as given below:
	\begin{equation*}
	\mathbf{C}_1 =
	\begin{bmatrix}
	\star & & & & & & \star & & & & &\\
	&\star & & & & & &\star & & & &\\
	& & \star & & & & & & \star & & &\\
	&  & & \star & & & &  & & \star & &\\
	&  & &  &\star & & &  & &  &\star &\\
	&  & &  & &\star & &  & &  & &\star
	\end{bmatrix}.
	\end{equation*}
	The columns of $\mathbf{C}_1$ are indexed from $(1,1)$ to $(3,4)$ in the lexicographic order from left to right. In the corresponding delivery array $\mathbf{B}_1$, $\star$'s can be filled satisfying condition D1. Then, each column of $\mathbf{B}_1$ contains four $\star$'s, and each row of $\mathbf{B}_1$  contains eight $\star$'s. In this case, we have $S=L/t=2$. The obtained delivery array $\mathbf{B}_1$ is given below.
	\begin{equation*}
	\mathbf{B}_1 =
	\begin{bmatrix}
	\star & 1  & \star & \star & 1  & \star & \star & 1 & \star & \star & 1 & \star\\
	\star & \star  & 1 & \star & \star  & 1 & \star & \star & 1 & \star & \star & 1\\
	1 & \star  & \star & 1 & \star  & \star & 1 & \star  & \star& 1 & \star  & \star\\
	\star & 2  & \star & \star & 2  & \star & \star & 2 & \star & \star & 2 & \star\\
	\star & \star  & 2 & \star & \star  & 2 & \star & \star & 2 & \star & \star & 2\\
	2 & \star  & \star & 2 & \star  & \star & 2 & \star  & \star& 2 & \star  & \star
	\end{bmatrix}.
	\end{equation*}
	The $(3,4,2,4,M,N)$  2D MACC scheme resulting from the above $(3,4,6,1)$ caching array $\mathbf{C}_1$ and $(\mathbf{C}_1,2,4,4)$ delivery array $\mathbf{B}_1$ has $M/N=1/6$

	The NDT of the corresponding $(3,4,2,4,M,N)$ MACC scheme is $S/F=1/3$. Also, the lower bound on the NDT is $(K_1K_2-r^2t)/(L+r^2t)=1/3$. Therefore, the NDT achieved is optimal under uncoded placement and one-shot delivery.
\end{exmp}

\subsection{ Comparison with the single-antenna scheme in \cite{ZWCC}}
  In \cite{ZWCC}, a few 2D MACC schemes are proposed, but all restricted to a single-antenna setting.
  Note that there are no multi-antenna schemes known for the 2D MACC setting.
   The best NDT given in \cite{ZWCC} is $K_1K_2(1-r^2M/N)/(1+K_1K_2M/N)$, when $r|K_1$ and $r|K_2$. Under the same constraint with $L$ transmit antennas, we achieved an NDT $K_1K_2(1-r^2M/N)/(L+K_1K_2M/N)$. The performance improvement is clearly because of the addition of the transmit antennas. It is also important to note that we could provide an instance of optimality of the 2D MACC scheme with multiple transmit antennas.  
 

\section{Conclusion}
\label{Conclusion}
In this work, we introduced the multi-antenna coded caching problem in a 2D MACC network. We presented coding schemes by constructing two special arrays, namely a caching array and a delivery array. An optimal MACC scheme is presented for certain parameter settings. Future work includes deriving converse bounds, extending optimality to a wider setting, and performance analysis with different precoders in low and moderate SNR regimes.
\section*{Acknowledgment}
This work was supported partly by the Science and Engineering Research Board (SERB) of Department of Science and Technology (DST), Government of India, through J.C. Bose National Fellowship to Prof. B. Sundar Rajan, and by the Ministry of Human Resource Development (MHRD), Government of India, through Prime Minister’s Research Fellowship (PMRF) to K. K. Krishnan Namboodiri and Elizabath Peter.

\newpage 

\end{document}